\documentclass{article}

\usepackage{microtype}
\usepackage{graphicx}
\usepackage{subcaption}
\usepackage{booktabs} % for professional tables

\usepackage{hyperref}

\usepackage[margin=1in]{geometry}
\usepackage{natbib}
\usepackage{doi}

\usepackage{amsmath}
\usepackage{amssymb}
\usepackage{mathtools}
\usepackage{amsthm}

\usepackage[capitalize,noabbrev]{cleveref}

\theoremstyle{plain}

\theoremstyle{definition}

\theoremstyle{remark}

\usepackage[textsize=tiny]{todonotes}

\usepackage{xspace}
\usepackage{booktabs,makecell}
\newcommand{\method}{AudioGuard\xspace}
\newcommand{\dataset}{AudioSafetyBench\xspace}

\usepackage{booktabs}
\usepackage{makecell}
\usepackage{xcolor}
\usepackage{pifont}
\definecolor{GoodGreen}{RGB}{34,139,34}
\definecolor{BadRed}{RGB}{178,34,34}
\usepackage{xcolor}
\usepackage{pifont}

\newcommand{\cmark}{\textcolor{green!70!black}{\scalebox{1.2}{\ding{51}}}} % 1.4 可调
\newcommand{\xmark}{\textcolor{red}{\scalebox{1.2}{\ding{55}}}}

\usepackage{wrapfig}
\usepackage{tcolorbox}
\usepackage{enumitem}
\usepackage{multirow}
\usepackage{xspace}
\usepackage{titletoc}
\usepackage{fontawesome}
\usepackage{tabularx}
\usepackage{array}
\newcolumntype{C}[1]{>{\centering\arraybackslash}m{#1}}
\newcolumntype{L}[1]{>{\raggedright\arraybackslash}m{#1}}

\usepackage{csquotes}

\usepackage{tcolorbox}
\newtcolorbox{prompt}[2][]{
  colback=yellow!10!white,   % Background color
  colframe=yellow!70!black,  % Frame color
  coltext=black,             % Text color
  fonttitle=\bfseries,       % Title font
  title=#2,                  % Title content
  #1,                        % Additional options
  fontupper=\ttfamily\small  % Monospaced + smaller text inside box
}

\usepackage{booktabs,tabularx,makecell}

\title{\method: Toward Comprehensive Audio Safety Protection Across Diverse Threat Models}
\author{
Mintong Kang \quad Chen Fang \quad Bo Li\\
University of Illinois Urbana-Champaign\\
\texttt{(mintong2, chenf2, lbo)@illinois.edu}
}
\date{}

\begin{document}
\maketitle

\begin{abstract}
Audio has rapidly become a primary interface for foundation models, powering real-time voice assistants. Ensuring safety in audio systems is inherently more complex than just ``unsafe text spoken aloud'': real-world risks can hinge on audio-native harmful sound events, speaker attributes (e.g., child voice), impersonation/voice-cloning misuse, and voice--content compositional harms, such as children voice plus sexual content. 
Such nature of audios makes it challenging to develop comprehensive benchmarks or guardrails against the unique risk landscape.
To close this gap, we conduct large-scale red teaming on audio systems, systematically uncover vulnerabilities in audios, and develop a comprehensive, policy-grounded audio risk taxonomy and generate \textbf{\dataset}, the first policy-based audio safety benchmark across diverse threat models. \dataset supports diverse languages, suspicious voices (e.g., celebrity/impersonation and child voice), risky voice and content combinations, and non-speech sound events. To defend against these threats, we propose \method, a unified guardrail consisting of 1) \textit{SoundGuard} for waveform-level audio-native detection and 2) \textit{ContentGuard} for policy-grounded semantic protection. Extensive experiments on \dataset and four complementary benchmarks show that \method consistently improves guardrail accuracy over strong audio-LLM-based baselines with substantially lower latency. 
% Together, our taxonomy, benchmark, and guardrail establish a unified foundation for comprehensive evaluation and interpretable audio safety in real-world deployments.
\end{abstract}

% \bo{
% 1. audio is important
% 2. challenges, thread model for audio is more complex, such as xxxxx, therefore there is no good benchmark, guardrail for these diverse audio risks, which could be severe.
% 3. to compose challenging benchmark considering these diverse threat models and real world vulnerabilities (enviroment)
% 4. to defend againsts such audio threat, we also propose guard xxxxx, composed of two parts xxxx.
% 5. results -- findings
% }

\section{Introduction}
\label{sec:intro}

As foundation models are increasingly deployed in consumer and enterprise products, audio has emerged as a primary interaction modality—powering real-time voice conversations with assistants (ChatGPT, Gemini) and speech-driven agent interfaces.
In parallel, the rapid maturation of \emph{text-to-speech (TTS)} and \emph{voice generation / voice cloning} services has made high-quality speech synthesis easy to integrate via APIs, including commercial platforms such as ElevenLabs~\citep{elevenlabs_tts_docs}, PlayHT~\citep{playht_api_docs}, and Resemble AI~\citep{resemble_docs}, as well as widely used cloud offerings like Amazon Polly~\citep{aws_polly_api_reference}, Microsoft Azure Speech TTS~\citep{azure_tts_docs}, and Google Cloud TTS~\citep{gcp_tts_docs}.
Yet ensuring safety in these settings is inherently more complex than ``unsafe text spoken aloud'': failures can hinge on audio-native harmful sound events, speaker attributes (e.g., child voice), impersonation/voice-cloning misuse, and voice--content compositional harms. As a result, the community still lacks standardized audio risk taxonomies, benchmarks, and practical guardrails that capture real-world audio safety.

To close this gap, we conduct large-scale red teaming of these audio-capable AI systems, uncovering systematic vulnerabilities specific to the audio modality. These failures span (i) \emph{audio-native} risks such as non-speech harmful sound events (e.g., sexual moans, human distressing), (ii) \emph{voice--content compositional} risks (e.g., child voice combined with sexual content), (iii) \emph{voice impersonation} and voice-cloning misuse, and (iv) transcript-level policy violations exacerbated by multilingual and diverse real-world scenarios.
Based on the red teaming practices, we develop a comprehensive, policy-grounded risk taxonomy and curate a standardized audio safety benchmark across language, celebrity, child voice, and audio application scenarios.
This process yields \textbf{\dataset}, the first comprehensive benchmark for \emph{audio-input} and \emph{audio-output} safety, enabling comprehensive measurement and fine-grained diagnosis of audio safety across realistic deployment settings and threat models.

To ensure audio safety across diverse scenarios and threat models, \textbf{guardrail models} provide a flexible and practical layer that can be deployed alongside heterogeneous audio systems with minimal impact on their core utility. 
However, existing audio guardrails often rely on large audio-language models as monolithic judges, which can be costly to run and brittle in practice, and they frequently inherit text-centric safety formulations that do not align with audio-specific risk taxonomies (e.g., non-speech sound events and voice--content compositional risks). 
To address these gaps, we propose \method, a unified audio guardrail that decouples audio-native cue detection from semantic safety guard: \textit{SoundGuard} extracts audio-native risk signals such as unsafe sound events and speaker identities directly from waveforms, \textit{ContentGuard} performs automatic speech recognition (ASR) followed by \textit{TextGuard} for guardrail over transcripts, and finally integrates both to produce interpretable, scenario-specific guardrail decisions.

Extensive experiments on \dataset and five complementary audio safety benchmarks show that \method consistently outperforms strong audio-LLM-based guardrails including GPT-Audio and Gemini 3 across diverse risk categories and application scenarios, while incurring substantially lower guardrail latency. We further validate the contributions of each component, demonstrating that the \textit{SoundGuard} and \textit{TextGuard} models in \method achieve higher accuracy than their respective counterparts and exhibit better alignment with audio-specific safety risks. Finally, we uncover practical training insights, including that fine-tuning on a single language can generalize to improved safety performance in other languages, yielding stronger cross-lingual guardrail behavior. Together, these results underscore the importance of jointly leveraging audio-native and semantic signals to achieve effective and interpretable audio safety in realistic deployments.

 We summarize the main \textbf{contributions} below:
\vspace{-0.25em}
\begin{itemize}\setlength{\itemsep}{0.15em}\setlength{\parskip}{0pt}\setlength{\parsep}{0pt}
    \item \textbf{Red-teaming driven audio risk discovery.} We conduct large-scale red teaming on audio-capable AI systems and voice generation pipelines, and develop a comprehensive, policy-grounded audio risk taxonomy that captures audio-native sound events, voice--content compositional risks, impersonation/voice cloning misuse, and multilingual scenario-specific violations.
    \item \textbf{Audio safety benchmark.} We develop \textbf{\dataset}, a comprehensive benchmark considering diverse threat models against \emph{audio-input} and \emph{audio-output} safety, built upon policy-grounded taxonomy and equipped with sliceable metadata across language, celebrity, child voice, and application scenarios.
    \item \textbf{Unified audio guardrail.} We develop \method, a flexible guardrail that combines \textit{SoundGuard} (audio-native cue detection), \textit{ContentGuard} (ASR + \textit{TextGuard}), and a compositional mechanism for scenario-specific guardrail across threat models.
    \item \textbf{Comprehensive evaluation and findings.} We evaluate \method on \dataset and five additional benchmarks, demonstrating consistent accuracy gains with lower latency (15\% guardrail accuracy gain while nearly half of latency compared to SOTA guardrail Gemini 3), strong component-wise improvements, and new gaurdrail training insights. 
    % \bo{any findings or numbers can mention here?}
\end{itemize}
\vspace{-0.5em}

\section{Related Work}

Recent progress in \textbf{large audio-language models (LALMs)} extends foundation models beyond text to process speech, general sound events, and music, enabling applications such as voice chat, audio question answering, and multi-turn spoken dialogue. Representative systems include SpeechGPT~\citep{zhang2023speechgpt}, SALMONN~\citep{tang2023salmonn}, Qwen2-Audio~\citep{chu2024qwen2audio}, and Audio Flamingo 3~\citep{goel2025audioflamingo3}. In parallel, the community has developed standardized evaluations for audio-centric instruction following and dialogue understanding, such as AIR-Bench~\citep{yang2024airbench} and ADU-Bench~\citep{gao2024adubench}. While these efforts rapidly expand the capabilities and deployment surface of audio-enabled AI, they do not yet provide a unified, policy-grounded framework for defining and measuring audio safety risks across realistic threat models.

In addition to audio understanding, modern speech generation has advanced rapidly, making high-fidelity \textbf{text-to-speech} and \textbf{voice cloning} widely accessible. Neural codec language models such as VALL-E~\citep{wang2023valle} and its cross-lingual extension VALL-E X~\citep{zhang2023vallex}, as well as large-scale generative speech models like Voicebox~\citep{le2023voicebox}, demonstrate strong in-context speaker conditioning, editing, and multilingual synthesis. These advances motivate safety evaluation beyond transcript-level policy violations to include voice impersonation, speaker-attribute risks (e.g., child voice), and voice--content compositional harms.

\textbf{Safety datasets and audio safety evaluation.}
Safety datasets such as DecodingTrust~\citep{wang2023decodingtrust}, HarmBench~\citep{mazeika2024harmbench}, AdvBench~\citep{zou2023universal}, ToxicChat~\citep{lin2023toxicchat}, BeaverTails~\citep{ji2024beavertails}, HarmfulQA~\citep{bhardwaj2023red}, JailbreakBench~\citep{chao2024jailbreakbench}, GuardBench~\citep{bassani2024guardbench}, and Poly-Guard~\citep{kang2025polyguard} are predominantly \emph{text-only}. Simply converting these prompts into speech (via TTS) does not meaningfully introduce \emph{audio-native} safety challenges, and thus fails to stress-test audio safety guardrails beyond transcript-level moderation. Several recent benchmarks begin to study safety in audio settings, including Omni-SafetyBench~\citep{pan2025omnisafetybenchbenchmarksafetyevaluation}, Jailbreak-AudioBench~\citep{cheng2025jailbreakaudiobenchindepthevaluationanalysis}, AdvWave~\citep{kang2024advwave}, AudioTrust~\citep{li2025audiotrust}, and AJailBench from Audio Jailbreak~\citep{song2025audiojailbreak}. These works are valuable and partially overlapping, but they differ in emphasis. In particular, AudioTrust studies broad trustworthiness dimensions of audio LLMs, whereas our work focuses more narrowly on \emph{policy-grounded audio-input/output safety guardrailing} with a deployable modular defense. Likewise, jailbreak-focused evaluations stress adversarial robustness, but do not by themselves provide a unified benchmark for audio-native, speaker-aware, compositional, and hard-benign policy evaluation. In contrast, \dataset is constructed from large-scale red teaming to enable comprehensive \emph{audio-input} and \emph{audio-output} safety evaluation, with a policy-grounded taxonomy and a standardized protocol that supports fine-grained slicing across language, celebrity, child voice, and diverse audio application scenarios.

\textbf{Audio attacks and robustness.}
Recent work has also highlighted the vulnerability of audio-capable systems to adversarial and jailbreak attacks. SpeechGuard~\citep{peri2024speechguard} studies adversarial robustness for multimodal models under speech perturbations, while AudioJailbreak~\citep{chen2025audiojailbreak} and Audio Jailbreak~\citep{song2025audiojailbreak} demonstrate increasingly realistic jailbreak attacks against large audio-language models. These results motivate defenses that can explicitly separate audio-native cues from transcript-level semantics, rather than relying only on monolithic judging.

\textbf{Guardrail models.}
Guardrail models provide a flexible way to improve foundation-model safety without modifying the underlying model~\citep{inan2023llama,chi2024llama,zeng2024shieldgemma,li2024salad,han2024wildguard,ghosh2024aegis,padhi2024granite,chensafewatch}. However, most existing guardrails are \emph{text-only} and therefore overlook audio-specific risks (e.g., non-speech harmful sound events, speaker attributes such as child voice, and voice impersonation or voice--content compositional harms). A straightforward alternative is to use large audio-language models as end-to-end safety judges (e.g., GPT-Audio, Gemini, Audio Flamingo~\citep{goel2025audioflamingo3}, Qwen3-Omni~\citep{xu2025qwen3omnitechnicalreport}), but such monolithic approaches can be expensive to deploy, sensitive to prompting, and—more importantly—often inherit text-centric safety formulations that do not align with audio-native risk taxonomies. Recent multimodal guardrail stacks such as Protect~\citep{avinash2025protect} move toward broader multimodal moderation, but are not specialized for the policy-grounded audio risk structure studied here. In contrast, \method offers a unified and efficient audio safety guardrail by decomposing the task into complementary components: \textit{SoundGuard} performs waveform-level audio-native cue detection, \textit{ContentGuard} applies \textit{ASR} followed by \textit{TextGuard} for policy-grounded semantic moderation, and an interpretable integration module composes both signals to produce scenario-specific decisions across threat models.

\begin{wrapfigure}{r}{0.42\textwidth}
    \centering
    \vspace{-5em}
    \includegraphics[width=0.42\textwidth]{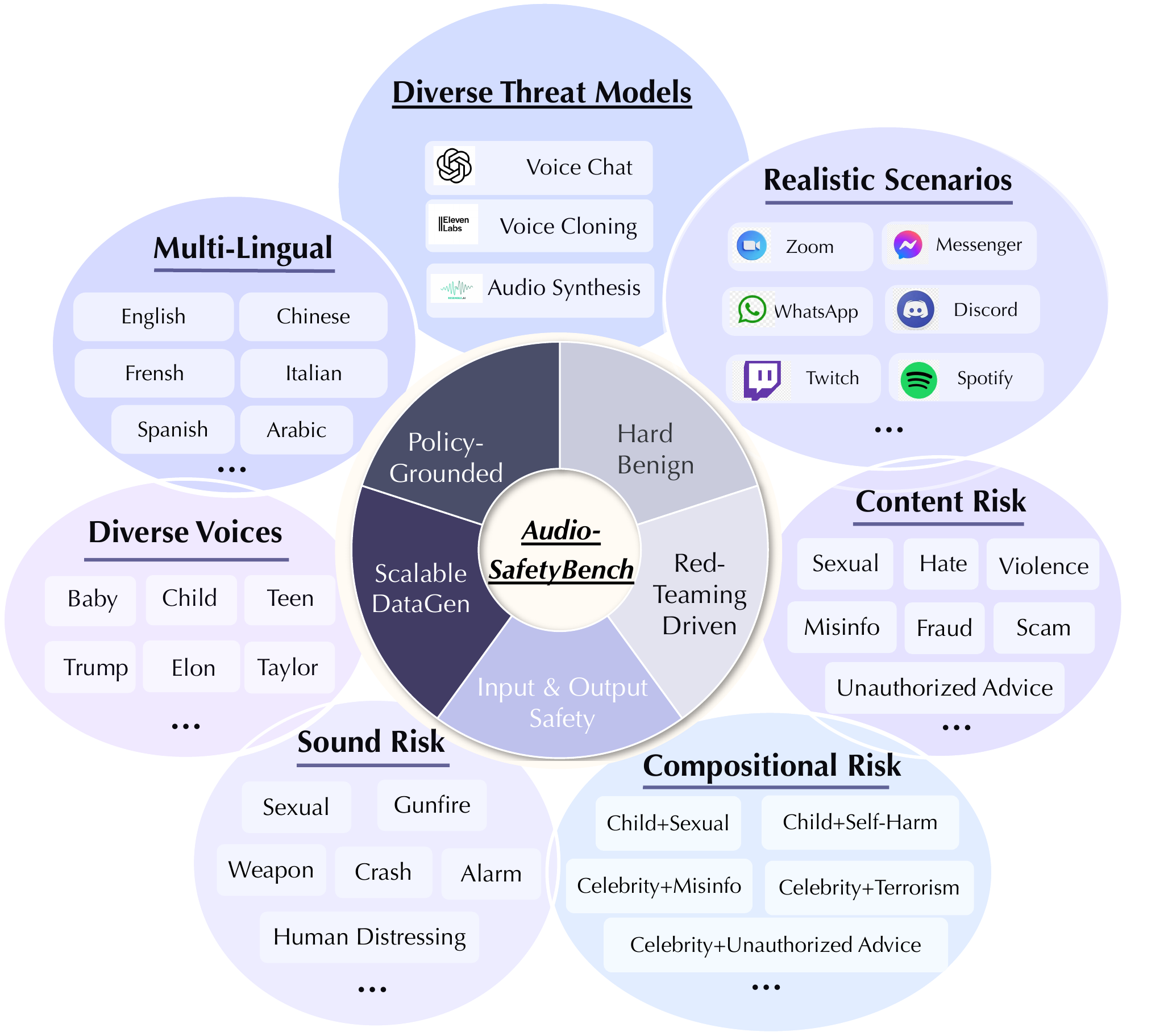}
    \caption{\small Overview of \dataset.}
    \label{fig:dataset}
    \vspace{-3.0em}
\end{wrapfigure}

\section{\dataset}
\label{sec:dataset}

In this section, we introduce \dataset, a comprehensive audio safety benchmark under realistic deployment settings. We first presents the motivation for audio-centered safety evaluation in \cref{subsec:overview}, then describe the end-to-end dataset development pipeline for \dataset in \cref{subsec:develop}, including taxonomy design, audio synthesis, red-teaming driven instance generation, hard-benign balancing, and multilingual augmentation. Finally, we summarize benchmark statistics and position \dataset against existing audio safety benchmarks in \cref{subsec:bench}. An overview of \dataset is provided in \Cref{fig:dataset}.

\subsection{Motivation}
\label{subsec:overview}

As audio-capable AI is increasingly used across diverse interaction scenarios (e.g., voice assistants, TTS/voice cloning, audio synthesis, and interactive voice agents), standardized audio safety evaluation must capture the key factors that drive real-world failures: \emph{audio-native} cues (e.g., non-speech harmful sound events), \emph{speaker-aware} risks (e.g., child voice and impersonation), and \emph{scenario-dependent} policies that differ across audio-input moderation and audio-output generation. Existing audio safety benchmarks typically cover a single threat model or primarily convert text-centric risks into speech, limiting their ability to diagnose these audio-specific vulnerabilities. \dataset is built to fill this gap by providing a policy-grounded taxonomy and a standardized benchmark for both \emph{audio-input} and \emph{audio-output} safety across diverse threat models.

\subsection{Development of \dataset}
\label{subsec:develop}

\textbf{Policy-grounded audio risk taxonomy from real-world scenarios.}
We first construct an audio-specific, policy-grounded risk taxonomy by distilling common safety requirements across realistic audio interaction scenarios. Concretely, we collect and consolidate publicly available safety policies from 20+ audio-related platforms spanning podcast sharing (e.g., Spotify), conferencing (e.g., Zoom), messaging/voice chat (e.g., WhatsApp, Discord), and live streaming (e.g., Twitch). We then parse these policy documents with an LLM-assisted pipeline to extract enforceable rules and map them into a unified, hierarchical taxonomy of audio risks.
Beyond transcript-level violations, the taxonomy explicitly includes: \textbf{(i) non-speech harmful sound events} (e.g., weapon handling cues, sexual sounds, distress screams), \textbf{(ii) impersonation and voice-cloning misuse} (e.g., restricted public-figure voices), and \textbf{(iii) voice--content compositional risks} where the audio unsafety depends jointly on voice and semantic content (e.g., child voice $\,+\,$ sexual content). This taxonomy provides a unified label space across applications and serves as the blueprint for benchmark instance construction and evaluation.

\textbf{Audio synthesis and control over voice conditions.}
To cover audio-native and speaker-aware risks beyond transcripts, we construct audio instances with explicit control over \emph{sound events}, \emph{speaker identity}, and \emph{spoken content}. 
For {non-speech harmful sound events}, we curate and scrape representative audio clips (e.g., weapon handling cues, explosion-like sounds, distress screams, sexual sounds) from public audio sources and sound-effect libraries, and optionally mix them with speech to create realistic mixed-signal contexts. 
For {impersonation and voice-cloning misuse}, we collect reference voice samples of public figures/celebrities, perform voice cloning to obtain a target voice, and synthesize policy-violation utterances via TTS under that cloned identity. 
For {speaker-attribute and compositional risks} (e.g., child voice $\,+\,$ sexual content), we similarly construct voice-conditioned audio by controlling speaker attributes (child vs.\ adult; target identity when applicable) and generating matched transcripts that realize the intended risk category. 
To make the spoken content better reflect real-world usage, we use LLMs to generate scenario-grounded utterances conditioned on safety policies and typical interaction patterns from the same 20+ platforms used for policy scraping, spanning podcast sharing, conferencing, messaging/voice chat, and live streaming.

\textbf{Red-teaming driven instance construction across diverse threat models.}
Guided by the taxonomy, we conduct large-scale red teaming to construct \emph{challenging} unsafe audio instances across multiple deployment-relevant threat models, including: (1) {audio-input moderation} for audio-language model I/O, (2) {audio-input moderation} for voice-cloning pipelines (screening user audio before cloning), (3) {audio-output safety} for TTS systems (screening synthesized speech before delivery), and (4) {interactive voice agents}, where both incoming user audio and outgoing agent audio must be monitored.
To ensure the resulting unsafe samples are \emph{non-trivial} (i.e., not easily rejected by default platform safeguards or detected by simple heuristics), we apply a multi-stage filtering procedure: we first generate candidate instances and then retain only those that (i) successfully pass basic TTS/voice-cloning generation constraints and produce natural, intelligible audio, (ii) are not flagged by off-the-shelf transcript-only moderation under benign transcription errors, and (iii) remain clearly policy-violating under the intended threat model when reviewed against our taxonomy by human check.

\textbf{Hard benign instances for false-positive stress testing.}
Audio safety systems must avoid over-blocking benign content that contains ambiguous cues (e.g., sirens in movies or sensitive keywords in educational contexts). To stress-test false positives, we curate a substantial set of \emph{hard benign} instances that are semantically safe but contain potentially triggering audio cues or keywords. These instances are constructed to mirror realistic benign uses (e.g., news reporting, historical discussion, safety training, entertainment audio) and are balanced against unsafe cases to support robust evaluation of both safety and utility.

\textbf{Multilingual augmentation.}
To reflect global deployment, we augment speech-involved audio with multilingual coverage by constructing parallel instances across \emph{17} languages. For each instance, we preserve the underlying risk intent and threat-model scenario while varying the linguistic realization of the spoken content. Concretely, we translate the source transcript using the Google Translate API and re-synthesize the corresponding speech under the same voice condition (e.g., celebrity/child voice when applicable), yielding multilingual audio pairs with matched semantics and controllable speaker attributes.

\begin{table*}[t]
\centering
\caption{Comparison between \dataset and other audio safety benchmarks.}
\label{tab:audg_feature_compare}
\vspace{-0.5em}
\resizebox{\textwidth}{!}{
\begin{tabular}{l|cccccc}
\toprule
\textbf{Benchmark} &
\makecell{\textbf{Diverse}\\\textbf{Threat Models}} & \makecell{\textbf{Input \& Output}\\\textbf{Safety}} &
\makecell{\textbf{Audio-Specific}\\\textbf{Risk Taxonomy}} &
\makecell{\textbf{Non-Speech}\\\textbf{Sound Events}}&
\makecell{\textbf{Diverse}\\\textbf{Voices}}&
\makecell{\textbf{Multi-}\\\textbf{Lingual}}  \\
\midrule
Nemotron-Content-Safety-Audio & \xmark & \xmark & \xmark & \xmark & \cmark & \xmark  \\
Jailbreak-AudioBench          & \xmark & \xmark & \xmark & \xmark & \cmark & \xmark \\
Omni-SafetyBench              & \xmark & \xmark & \xmark & \xmark & \xmark & \xmark \\
AdvWave                       & \xmark & \xmark & \xmark & \cmark & \xmark & \xmark \\
\dataset (Ours)               & \cmark & \cmark & \cmark & \cmark & \cmark & \cmark \\
\bottomrule
\end{tabular}
}
\vspace{-1.2em}
\end{table*}

\subsection{Analysis of \dataset}
\label{subsec:bench}
\dataset presents a large-scale benchmark for \emph{audio-input} and \emph{audio-output} safety, constructed via policy grounding and red-teaming driven instance generation. It contains \textbf{10K+} labeled audio instances spanning \textbf{17} languages and \textbf{50+} speaker identities, with rich, sliceable metadata covering diverse voice conditions (including celebrity/impersonation and child voice), non-speech harmful sound events, and multiple real-world audio application scenarios. The benchmark is also designed to support fine-grained diagnosis across risk categories, languages, speaker conditions, and audio provenance.

\cref{tab:audg_feature_compare} situates \dataset relative to existing audio safety benchmarks. Prior datasets typically focus on a single threat model or a narrow class of attacks, and often lack either (i) coverage of both input and output safety, (ii) an audio-specific risk taxonomy beyond transcript-level violations, (iii) explicit non-speech sound-event evaluation, (iv) diverse voice conditions, or (v) multilingual breadth. In contrast, \dataset jointly covers all of these dimensions, enabling comprehensive and realistic evaluation of audio safety systems.

\section{\method}
\label{sec:method}

In this section, we present \method, a unified audio safety guardrail that composes audio-native cue detection with transcript-based semantic moderation. We first provide an overview of the framework and its design principles in \cref{subsec:method_overview}, then detail the model architecture, including \textit{SoundGuard}, \textit{ContentGuard} (ASR+\textit{TextGuard}), and the compositional integration in \cref{subsec:method_arch}. Finally, we describe the training and configuration for each component and the overall system in \cref{subsec:method_training}.

\subsection{Motivation and Overview}
\label{subsec:method_overview}

To safeguard audio safety under diverse threat models, guardrails offer a flexible layer that can be deployed alongside heterogeneous AI systems with minimal impact on benign utility.
To develop an audio safety guardrail, a straightforward approach is to fine-tune a single audio-language model as a monolithic judge. In practice, this design is difficult to train and deploy: (1) it requires large-scale audio supervision that jointly annotates {audio-native} cues (e.g., non-speech sound events, speaker attributes) and {semantic} policy violations, which is expensive and hard to curate at scale; and (2) it is inflexible across heterogeneous threat models. For example, in voice-cloning pipelines the primary requirement may be to detect restricted speaker identities, whereas in voice chat the dominant risks are transcript-level policy violations; forcing a single model to cover all cases often depends on prompt-based task re-framing, which can introduce distribution shifts and brittle performance.

\begin{figure*}[t]
  \centering
  \includegraphics[width=0.75\textwidth]{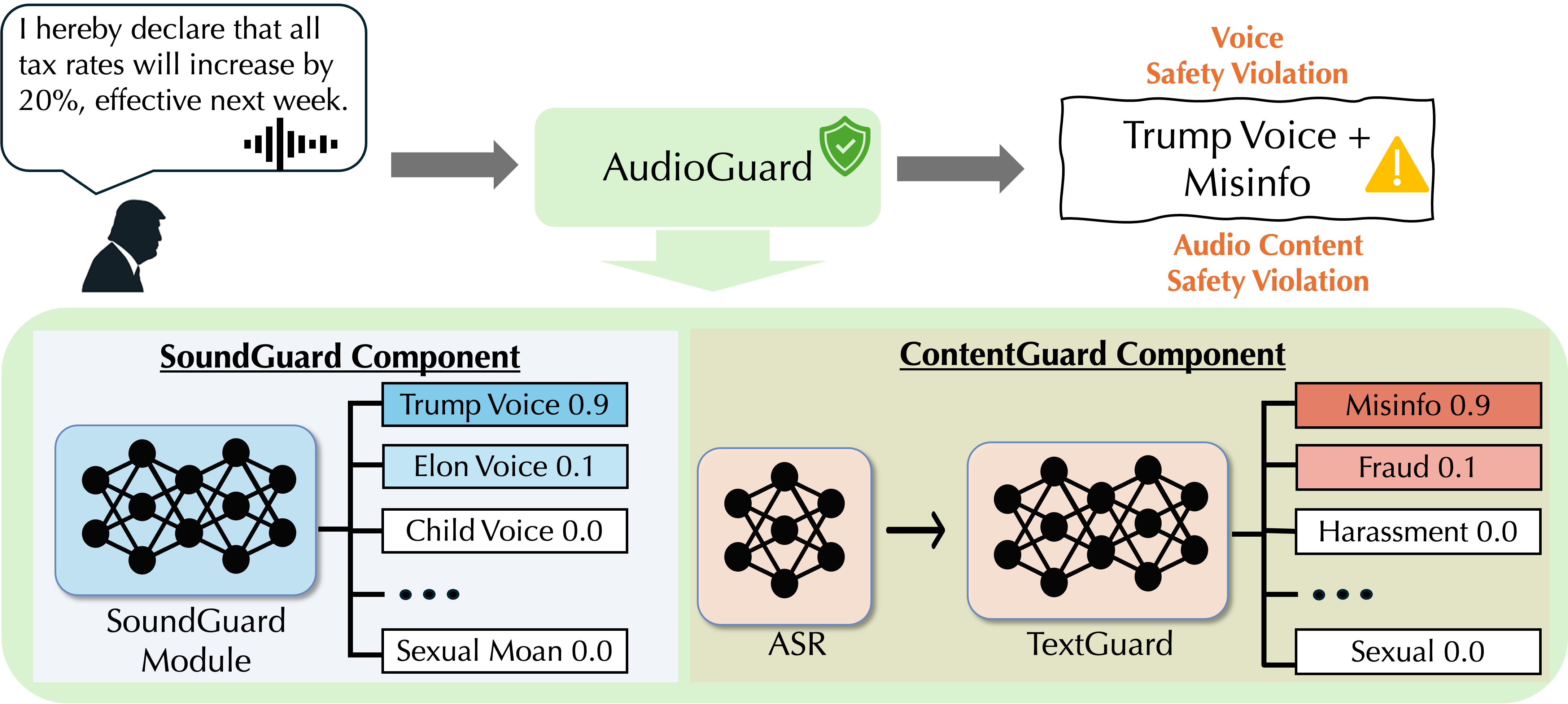}
  % \vspace{-0.5em}
  \caption{\textbf{AudioGuard framework overview.} Given an input audio waveform, \textit{SoundGuard} detects audio-native safety cues directly from the signal (e.g., speaker identity, child voice, gunshot, sexual sounds) and outputs sound risk scores. In parallel, \textit{ContentGuard} transcribes the audio via ASR and leverages \textit{TextGuard} to predict the content risk scores (e.g., misinformation, fraud, harassment, sexual content). \method provides the end-to-end audio guardrail with interpretable predictions under diverse audio threat models.}
  \label{fig:audioguard-framework}
  % \vspace{-1em}
\end{figure*}

To address these challenges, we propose \method, a dual-path audio safety guardrail that explicitly decomposes safety reasoning into \emph{audio-native} and \emph{semantic} channels and then composes them into a scenario-specific decision (Figure~\ref{fig:audioguard-framework}). Given an input audio waveform $x$, \method produces (i) \textbf{sound/voice risk scores} via \textit{SoundGuard}, (ii) \textbf{content risk scores} via \textit{ContentGuard} (\textit{automatic speech recognition (ASR)} $\rightarrow$ \textit{TextGuard}), and (iii) a final guardrail action for the target threat model (e.g., allow or block). This decomposition directly targets two practical failure modes: (1) critical risks are often \emph{not} recoverable from transcripts alone (e.g., non-speech harmful sound events or speaker identity/child voice), and (2) real policies are frequently \emph{compositional} and scenario-dependent (e.g., \emph{public-figure voice} $\,+\,$ \emph{misinformation}), requiring joint reasoning over both audio-native and semantic signals.

\subsection{\method Model}
\label{subsec:method_arch}

We detail the three components of \method: \textit{SoundGuard} for audio-native cue detection, \textit{ContentGuard} for transcript-based semantic moderation, and a compositional integration module that maps both signals to scenario-specific guardrail actions.

\textbf{SoundGuard: audio-native cue detection.}
\textit{SoundGuard} takes an input waveform $x$ and predicts audio-native cues that are directly observable from the signal, including \emph{speaker attributes} (e.g., child voice, celebrity) and \emph{non-speech sound events} (e.g., gunfire/explosion-like sounds, distress screams, sexual sounds). Formally, SoundGuard outputs a multi-label score vector
\begin{equation}
\mathbf{s} = \mathrm{SoundGuard}(x) \in [0,1]^{K_s},
\end{equation}
where each dimension corresponds to a sound/voice cue in our taxonomy (or a calibrated grouping of cues), and $s_k$ denotes the confidence of cue $k$.

\textbf{ContentGuard: ASR + transcript-based safety reasoning.}
\textit{ContentGuard} targets policy-grounded semantic violations expressed in speech. It first transcribes audio with an ASR model
\begin{equation}
\hat{t} = \mathrm{ASR}(x),
\end{equation}
and then applies \textit{TextGuard} to the transcript to predict content risk scores
\begin{equation}
\mathbf{c} = \mathrm{TextGuard}(\hat{t}) \in [0,1]^{K_c},
\end{equation}
where each dimension corresponds to a policy-grounded content risk category (e.g., fraud, harassment, sexual content, misinformation). In practice, TextGuard can be instantiated as a lightweight classifier or an instruction-tuned LM fine-tuned to output structured risk labels/scores.

\textbf{Compositional integration for scenario-specific decisions.}
\method combines $\mathbf{s}$ and $\mathbf{c}$ into a unified, interpretable guardrail decision via compositional integration. The integration is configured per threat model to reflect scenario-specific constraints (e.g., voice cloning/TTS vs.\ voice chat vs.\ voice agents). We represent each policy rule $r$ as a conjunction of threshold tests over sound/voice cues and content risks:
\begin{equation}
\phi_r(\mathbf{s},\mathbf{c}) =
\Big(\bigwedge_{k \in \mathcal{S}_r} s_k \ge \tau_k\Big)\ \wedge\
\Big(\bigwedge_{\ell \in \mathcal{C}_r} c_\ell \ge \tau_\ell\Big),
\end{equation}
where $\mathcal{S}_r \subseteq \{1,\dots,K_s\}$ indexes the subset of SoundGuard cues used by rule $r$ (e.g., \emph{public-figure voice}, \emph{child voice}, \emph{gunfire}), $\mathcal{C}_r \subseteq \{1,\dots,K_c\}$ indexes the subset of TextGuard content risks used by rule $r$ (e.g., \emph{misinformation}, \emph{sexual content}), and $\{\tau_k\}, \{\tau_\ell\}$ are rule-specific thresholds. Triggered rules are mapped to actions (e.g., \textsc{Allow}, \textsc{Block}, \textsc{Review}) via a priority-ordered rule list.
This decomposition provides: (i) \textbf{interpretability}, since decisions are attributable to specific triggered cues/risks; and (ii) \textbf{flexibility}, since policies can be updated or specialized to new threat models by changing rule definitions and thresholds without retraining SoundGuard or TextGuard.

\textbf{Implementation details and label spaces.}
For reproducibility, we instantiate {ContentGuard} with \textbf{Whisper-Large-v3} as the ASR model. {TextGuard} predicts policy-grounded semantic risk labels from the transcript, while {SoundGuard} predicts speaker-aware and audio-native cues from the waveform. We provide the full \textit{TextGuard} label inventory and the \textit{SoundGuard} label-space schema in Appendix~\ref{app:exp}.

\subsection{Training of \method}
\label{subsec:method_training}

We next describe how we train the two learnable components of \method (SoundGuard and TextGuard), and how we configure the model for efficiency and flexibility.

\textbf{Training SoundGuard.}
SoundGuard is implemented as a lightweight audio classifier on top of a pretrained encoder. Concretely, we adopt the SpeechBrain ECAPA-TDNN speaker encoder (\path{speechbrain/spkrec-ecapa-voxceleb}) as the default backbone feature extractor and attach an MLP prediction head for multi-label cue classification. During training, we keep the backbone fixed (unless otherwise noted) and fine-tune the MLP head (and optionally the last encoder block) to predict audio-native cues. We train SoundGuard as a multi-label predictor over audio-native cues using a mixture of (i) curated non-speech sound-event clips and mixtures (speech$+$event), and (ii) voice-conditioned data for speaker. Given cue labels $\mathbf{y}^{(s)} \in \{0,1\}^{K_s}$, we optimize the binary cross-entropy objective:
\begin{equation}
\mathcal{L}_{\text{sound}} = \sum_{k=1}^{K_s} \mathrm{BCE}(s_k, y^{(s)}_k).
\end{equation}

\textbf{Training TextGuard.}
TextGuard is implemented by fine-tuning a lightweight instruction-tuned LLM (Gemma-3-it) for policy-grounded transcript moderation. Starting from a pretrained Gemma-3 checkpoint, we fine-tune the model to predict a \textsc{Safe} label or a fine-grained unsafe risk category from our taxonomy, given text input. Concretely, we attach a small classification head (linear layer) on top of the final hidden state and optimize cross-entropy over the target label space. Training supervision is derived from our policy-grounded taxonomy, and based on that we develop text safety data following the construction pipeline of Poly-Guard~\citep{kang2025polyguard}. Because ASR transcripts can exhibit distribution shift relative to clean text (e.g., recognition errors and punctuation loss), we improve robustness via transcript-noise augmentation. During training and evaluation, we inject realistic perturbations by passing text through a \emph{TTS$\rightarrow$ASR} round-trip to simulate recognition artifacts. This encourages TextGuard to remain stable under transcription noise and reduces brittleness to ASR-specific artifacts; we ablate this design in Appendix~\ref{app:exp}.

\textbf{End-to-end efficiency.}
By decoupling waveform-level cue detection from transcript-level semantic moderation, \method avoids relying on a large audio-LLM as a monolithic judge. The resulting guardrail is efficient, \textit{SoundGuard} and \textit{ContentGuard} can run in parallel, and modular, allowing components to be swapped or improved independently. Moreover, \method is flexible across threat models: deployments can enable only the necessary submodules (e.g., \textit{SoundGuard}-only for speaker-identity screening in voice cloning) to meet latency and cost constraints.

% =========================
% REPLACE: \section{Evaluation Result} with this section title + opening + setup + main results
% =========================
\section{Evaluation Results}
\label{sec:exp}

Through evaluations on five audio safety benchmarks, we find that \textbf{(1)} \method consistently outperforms strong audio-LLM guardrail baselines in overall joint sound$+$content accuracy while achieving lower end-to-end latency; \textbf{(2)} the largest gaps emerge on severe \emph{voice--content compositional} risks, especially on public-figure risks that pose major challenges for existing guardrails; \textbf{(3)} non-speech harmful sound events remain a major blind spot for transcript-centric guardrails and are handled more reliably by \method; and \textbf{(4)} TextGuard in \method exhibits strong cross-lingual transfer, where English-only training still yields consistent gains across 17 languages.

\subsection{Evaluation setup}

\textbf{Datasets.} We evaluate on \dataset and four external audio safety benchmarks. \dataset contains three splits designed to isolate key audio safety vulnerabilities: \textbf{Speech} (speech inputs with \emph{voice--content compositional} risks), \textbf{Non-Speech} (audio-native harmful sound events), and \textbf{ElevenLabs Red-Teaming} (realistic, in-the-wild voice-generation/red-teaming cases with speaker-aware and compositional risks). For external benchmarks, \textbf{Jailbreak-AudioBench}~\citep{cheng2025jailbreakaudiobenchindepthevaluationanalysis} evaluates audio jailbreak attacks; \textbf{Nemotron-Content-Safety-Audio}~\citep{ghosh-etal-2025-aegis2} provides a content-safety test set with speaker and violation metadata; \textbf{Omni-SafetyBench}~\citep{pan2025omnisafetybenchbenchmarksafetyevaluation} includes multimodal safety evaluations with an audio subset covering diverse attack types; and \textbf{AdvWave}~\citep{kang2024advwave} contains adversarial audio attacks generated via wave-based perturbations.

\textbf{Baselines.}
We compare \method against advanced audio-LLMs used as \emph{end-to-end} guardrails via prompting, including Gemma 3n~\citep{google_gemma3n_e2b_hf}, Qwen3-Omni~\citep{xu2025qwen3omnitechnicalreport}, Audio Flamingo 3~\citep{goel2025audioflamingo3}, Gemini 3, and GPT-Audio. For each audio-LLM baseline, we prompt the model to act as a safety judge given an input waveform and to output structured predictions aligned with our taxonomy. We use a fixed output schema and exact-string parsing; the full prompt template and parsing protocol are provided in Appendix~\ref{app:exp}. When an external benchmark does not align exactly with our 15-category taxonomy, we evaluate all methods consistently at the \textsc{safe} vs.\ \textsc{unsafe} level.

\textbf{Metric and latency protocol.}
For \cref{tab:main}, a prediction is counted as correct only when \emph{both} the \textbf{sound-risk} prediction and the \textbf{content-risk} prediction match the ground truth. For non-speech-only clips, correctness depends only on the audio-native risk label. We report end-to-end wall-clock latency per sample. \method is measured locally on an \textbf{NVIDIA A6000} GPU, while proprietary baselines such as Gemini 3 and GPT-Audio are measured using API end-to-end wall-clock time; thus the reported latency reflects deployment-oriented end-to-end latency rather than isolated model-only compute time.

\begin{table*}[t]
\centering
\caption{
\textbf{Overall audio safety guardrail accuracy across benchmarks.}
We evaluate on \dataset (speech, non-speech, and ElevenLabs red-teaming split) and four external audio safety benchmarks. 
A prediction is counted as correct only when \emph{both} the \textbf{sound-risk} (audio-native cues) and the \textbf{content-risk} (transcript-level safety prediction) match the ground truth.
We also report average performance across all benchmarks and end-to-end per-sample latency (seconds).
% \bo{we treat the the prediction to be accurate when the sound xxxx and content xxxxx}
% \bo{add latency column}
}
% \vspace{-0.5em}
\label{tab:main}
\resizebox{\textwidth}{!}{
\begin{tabular}{l|ccc|cccc|c|c}
\toprule
\multirow{4}{*}{\textbf{Method}} &
\multicolumn{3}{c|}{\textbf{\dataset}} &
\multirow{4}{*}{\makecell{\textbf{Nemotron-}\\~\textbf{Audio}}} &
\multirow{4}{*}{\makecell{\textbf{Jailbreak-}\\~\textbf{AudioBench}}} &
\multirow{4}{*}{\makecell{\textbf{Omni-}\\~\textbf{SafetyBench}}} &
\multirow{4}{*}{\makecell{\textbf{AdvWave}}} &
\multirow{4}{*}{\textbf{Avg}}&
\multirow{4}{*}{\textbf{Latency (s)}} \\
\cmidrule(lr){2-4}
& \makecell{\textbf{Speech}} &
\textbf{Non-Speech} &
\makecell{\textbf{ElevenLabs}\\~\textbf{Red-Teaming Test}} & & & & & \\
\midrule
Gemma 3         & 0.194 & 0.379 & 0.427 & 0.686 & 0.829 & 0.462 & 0.692 & 0.524 & 0.823 \\
Qwen3-Omni      & 0.245 & 0.607 & 0.362 & 0.723 & 0.682 & 0.536 & 0.872 & 0.575 & 1.834 \\
Audio Flamingo 3& 0.235 & 0.597 & 0.103 & 0.800 & 0.430 & 0.550 & 0.780 & 0.499 & 1.042 \\
Gemini 3        & 0.357 & 0.749 & 0.732 & 0.900 & 0.862 & 0.680 & 0.900 & 0.740 & 3.245 \\
GPT-Audio       & 0.289 & 0.720 & 0.400 & 0.894 & 0.829 & 0.652 & 0.918 & 0.672 & 2.542 \\
\midrule
\method         & \textbf{0.832} & \textbf{0.876} & \textbf{0.890} & \textbf{0.913} & \textbf{0.893} & \textbf{0.743} & \textbf{0.953} & \textbf{0.871} & 1.423 \\
\bottomrule
\end{tabular}
}
% \vspace{-1em}
\end{table*}

\subsection{Main Results}
\label{subsec:exp_main}

\Cref{tab:main} reports joint sound$+$content accuracy across \dataset and four external benchmarks. Across all settings, \method substantially outperforms end-to-end audio-LLM guardrails, improving the average accuracy from $0.740$ (Gemini 3) and $0.672$ (GPT-Audio) to $0.871$. The gains are most pronounced on \dataset, where baselines struggle with audio-native and compositional failures: on the \dataset \textbf{Speech} split, \method achieves $0.832$ versus $0.357$ (Gemini 3) and $0.289$ (GPT-Audio); on the \textbf{Non-Speech} split, \method reaches $0.876$ versus $0.749$ and $0.720$; and on the realistic \textbf{ElevenLabs Red-Teaming} split, \method attains $0.890$ versus $0.732$ and $0.400$. Notably, even on established external benchmarks where strong audio-LLMs perform well (e.g., Nemotron-Audio and Jailbreak-AudioBench), \method remains consistently best, indicating that the improvements are not limited to in-domain data.

Beyond accuracy, \method offers a favorable efficiency profile. While large audio-LLM judges can be slow (e.g., $3.245$s for Gemini 3 and $2.542$s for GPT-Audio), \method achieves higher accuracy with substantially lower latency ($1.423$s), enabled by parallelizable \textit{SoundGuard} and \textit{ContentGuard} and a lightweight compositional integration. Together, these results highlight that decomposing audio safety into complementary audio-native and semantic signals yields both stronger robustness and better deployment efficiency than monolithic audio-LLM guardrails.

\begin{figure*}[t]
  \centering
  \begin{subfigure}[t]{0.49\textwidth}
    \centering
    \includegraphics[width=\textwidth]{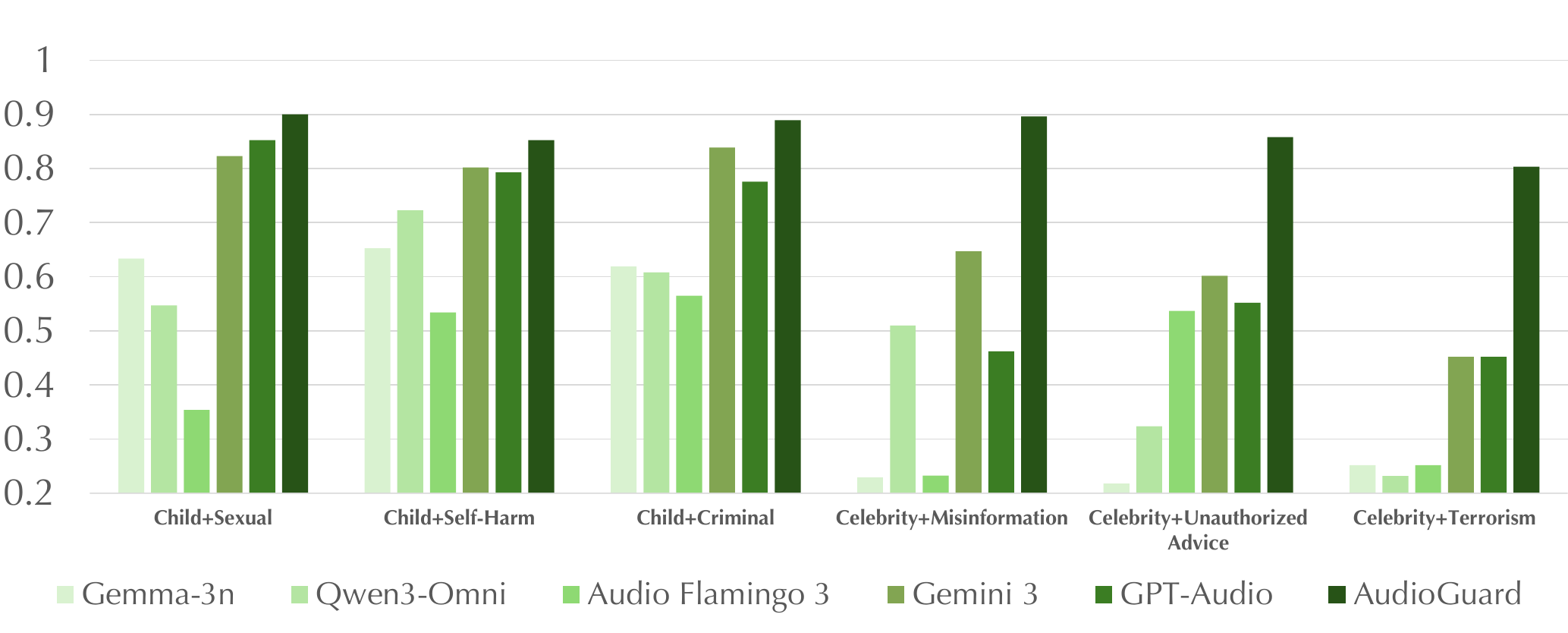}
    \caption{\scriptsize \dataset (Speech): severe voice--content compositional slices.}
    \label{fig:severe_combinations}
  \end{subfigure}\hfill
  \begin{subfigure}[t]{0.49\textwidth}
    \centering
    \includegraphics[width=\textwidth]{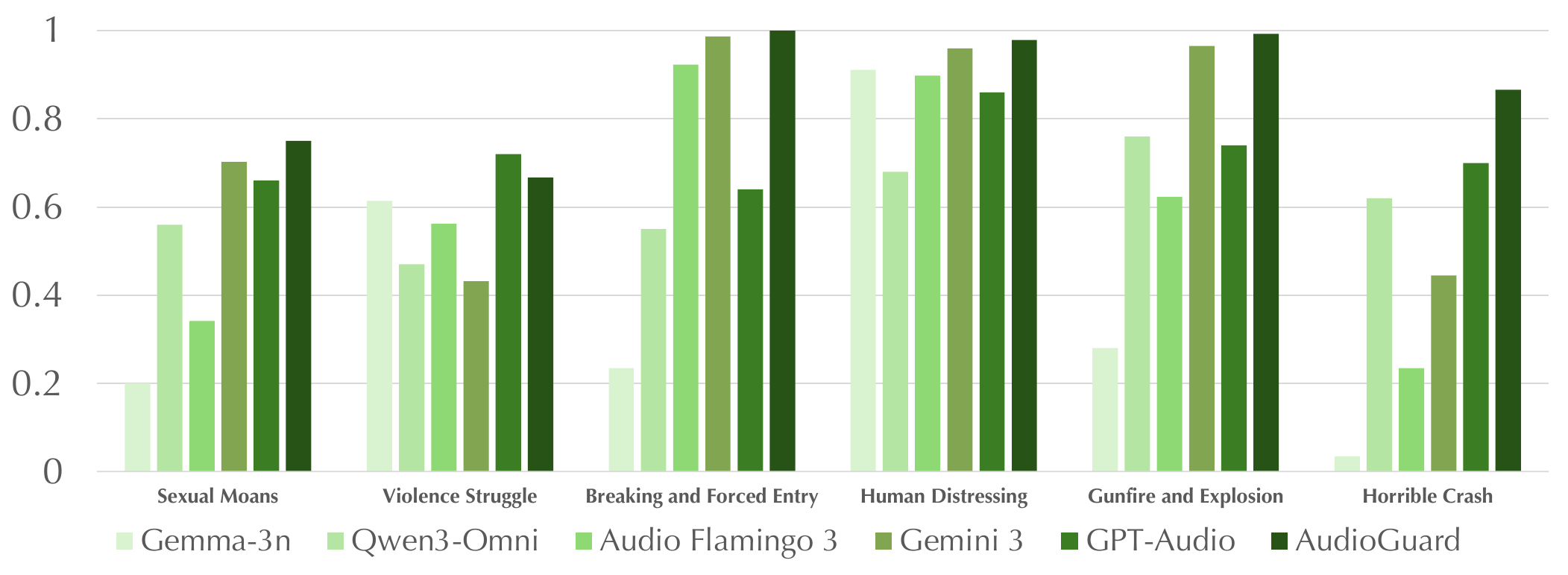}
    \caption{\scriptsize \dataset (Non-Speech): severe sound-event slices.}
    \label{fig:nonspeech_sound_events}
  \end{subfigure}
    \vspace{-0.5em}
  \caption{\textbf{Category-wise guardrail performance for audio-specific risks in \dataset.}
  \textbf{Left:} joint sound\,+\,content accuracy (higher the better) on representative severe \emph{voice--content compositional} risks, where a high-risk voice attribute (child voice or celebrity/impersonation) co-occurs with a semantic risk category (e.g., Sexual, Self-Harm, Criminal, Misinformation, Unauthorized Advice, Terrorism); a prediction is correct only if \emph{both} the voice/audio cue and transcript/content risk match the ground truth.
  \textbf{Right:} accuracy on representative \emph{non-speech} harmful sound events (e.g., sexual moans, violence/struggle, breaking/forced entry, human distress, gunfire/explosions, crash sounds); since no transcript is involved, correctness depends only on the predicted audio risk category.}
  \label{fig:slice_diagnosis}
  \vspace{-1em}
\end{figure*}

\subsection{Fine-grained diagnosis on audio risks}
\label{subsec:slice_analysis}

While overall results in \cref{tab:main} establish consistent gains, \dataset is designed to enable \emph{risk-category-wise diagnosis} of audio-specific vulnerabilities that are often obscured by aggregate metrics. Accordingly, Figures~\ref{fig:severe_combinations}--\ref{fig:nonspeech_sound_events} present two representative category-wise results: (i) severe \emph{voice--content compositional} harms in the Speech split, and (ii) \emph{non-speech} harmful sound-event slices where transcript-only reasoning is unavailable.

\textbf{Severe voice--content compositional vulnerabilities.}
\cref{fig:severe_combinations} reports joint sound$+$content accuracy on challenging cases where a high-risk \emph{voice attribute} (child voice or celebrity/impersonation) co-occurs with a semantic risk category (e.g., Sexual, Self-Harm, Criminal, Misinformation). Across these slices, end-to-end audio-LLM judges may correctly identify the semantic risk from the transcript yet often fail to reliably detect the voice attribute, leading to low \emph{joint} accuracy under compositional policies. The gap is especially apparent in celebrity/impersonation combinations (e.g., celebrity$+$misinformation or celebrity$+$terrorism), where correctly moderating requires \emph{simultaneous} speaker-aware detection and content moderation. In contrast, \method remains strong across all combinations, reflecting the benefit of explicitly separating waveform-level cue detection (\textit{SoundGuard}) from semantic moderation (\textit{ContentGuard}) and composing them via interpretable rules aligned with the threat model.

\textbf{Non-speech harmful sound events remain challenging for monolithic judges.}
\cref{fig:nonspeech_sound_events} evaluates representative non-speech sound-event slices. Since these clips contain no spoken content, transcript-based reasoning provides little to no signal. We observe that audio-LLM baselines can be brittle in this setting, likely due to limited calibration for acoustically diverse events and the absence of linguistic context. \method achieves consistently higher accuracy across event types, indicating that a dedicated waveform-level detector trained on audio-native cues is critical for capturing these risks.

% \begin{figure*}[h]
%   \centering
%   \includegraphics[width=0.8\textwidth]{figs/exp1.pdf}
%   \caption{\textbf{Guardrail accuracy on representative severe voice--content compositional slices in \dataset (Speech).} Accuracy (higher is better) on challenging combinations where a high-risk voice attribute (child voice or celebrity/impersonation) co-occurs with a semantic risk category (e.g., Sexual, Self-Harm, Criminal, Misinformation, Unauthorized Advice, Terrorism). We report joint sound\,+\,content accuracy, counting a prediction as correct only when \emph{both} the audio-native (voice) risk and transcript/content risk decisions match the ground truth.}
%   \label{fig:severe_combinations}
% \end{figure*}

% \begin{figure*}[h]
%   \centering
%   \includegraphics[width=0.92\textwidth]{figs/exp2.pdf}
%   \caption{\textbf{Guardrail accuracy on representative non-speech harmful sound-event slices in \dataset (Non-Speech).}
% Accuracy (higher is better) on challenging non-speech sound events (e.g., sexual moans, violence/struggle, breaking/forced entry, human distress, gunfire/explosions, crash sounds). Since there is no content/transcript involved, we report joint sound accuracy, counting a prediction as correct when the predicted audio risk label matches the ground truth.}
%   \label{fig:nonspeech_sound_events}
% \end{figure*}

\subsection{Ablation Studies}
\label{subsec:ablation}

We ablate \method by analyzing its two core components---\textit{TextGuard} for transcript-level semantic moderation and \textit{SoundGuard} for waveform-level cue detection---against the same set of audio-LLM baselines. We further study multilingual transfer, threshold sensitivity, backbone choice, transcript-noise augmentation, and noisy-overlap robustness in Appendix~\ref{app:exp}.

\textbf{TextGuard improves policy-grounded semantic moderation.}
\Cref{tab:text_guardrail_comparison} in Appendix~\ref{app:exp} reports \emph{content-only} accuracy, i.e., correctness of safe/unsafe content prediction from transcripts. \textit{TextGuard} achieves the strongest overall performance across benchmarks, with clear gains on \dataset (0.936 on Speech and 0.959 on ElevenLabs) and strong results on external benchmarks as well. These results suggest that the improvements of \method are not solely driven by better audio-native detection, but also by stronger transcript-side safety reasoning aligned with our policy-grounded taxonomy. We further verify in \cref{tab:tts_asr_aug} that TTS$\rightarrow$ASR augmentation materially improves transcript robustness, indicating that explicit training for ASR-style noise is important for deployment-time semantic moderation.

\textbf{SoundGuard is necessary for audio-native and speaker-aware risks.}
\Cref{tab:sound_guardrail_comparison} in Appendix~\ref{app:exp} reports sound-only accuracy on \dataset splits. \textit{SoundGuard} substantially outperforms all audio-LLM baselines on \dataset, improving accuracy from 0.357/0.289 (Gemini 3/GPT-Audio) to 0.832 on Speech, from 0.573/0.452 to 0.815 on Non-Speech, and from 0.732/0.400 to 0.790 on ElevenLabs red-teaming. These gains highlight that audio-native cues (e.g., non-speech sound events and speaker attributes) are difficult to recover reliably via transcript-centric reasoning or monolithic prompting, and instead benefit from a dedicated waveform-level detector trained on audio-native supervision. We also study the effect of the SoundGuard encoder in \cref{tab:stronger_encoder}, where a stronger backbone does not uniformly improve end-to-end guardrail quality, suggesting that robustness depends not only on encoder scale but also on how well the representation matches the audio safety task.

\textbf{Cross-lingual generalization of TextGuard.}
\Cref{tab:multilingual_f1} in Appendix~\ref{app:exp} shows that \textit{TextGuard} trained with \emph{English-only} supervision still generalizes strongly to non-English safety classification. Despite never seeing labeled training data in other languages, \textit{TextGuard} consistently improves F1 over the base Gemma-1B across all 17 languages, indicating that policy-grounded decision boundaries learned from English transfer effectively to multilingual settings. This suggests that, in practice, high-quality supervision in a single language can already yield meaningful multilingual robustness, reducing the annotation burden for deploying transcript-based guardrails globally.

\textbf{Integration and operating-point sensitivity.}
Beyond the component-wise ablations, we also examine the compositional integration in \cref{tab:threshold_sensitivity}. The best performance is achieved at an intermediate threshold setting, while overly permissive or overly conservative thresholds both degrade end-to-end accuracy. This result supports our rule-based integration design and shows that \method benefits from a stable, interpretable operating region rather than narrow threshold tuning.

\section{Conclusions}

We bridge the gap between real-world audio deployments and text-centric safety evaluation by introducing a policy-grounded audio risk taxonomy and \dataset, a standardized benchmark for audio-input and audio-output safety. We further propose \method, a modular guardrail that combines waveform-level audio-native cue detection with transcript-level semantic moderation. Across \dataset and five complementary benchmarks, \method improves joint accuracy and reduces latency versus monolithic audio-LLM judges, while enabling fine-grained analysis of compositional, non-speech, and multilingual failures.

\section*{Impact Statement}
This work aims to improve the safety of audio-capable AI systems by providing a standardized benchmark (\dataset) and an efficient, interpretable guardrail (\method) for both audio-input and audio-output settings. By explicitly modeling audio-native cues (e.g., non-speech harmful sound events and speaker attributes) and voice--content compositional risks, our approach can help developers better evaluate and mitigate safety failures in real-world voice assistants, TTS/voice-cloning services, and interactive voice agents. 

Potential positive impacts include reducing harmful or policy-violating audio generation, improving detection of impersonation and child-safety risks, and enabling more transparent, diagnosis-driven safety improvements through slice-based evaluation. Potential risks include dual-use concerns: the benchmark taxonomy and red-teaming insights could inform adversaries about failure modes, and voice-related evaluation may be misused to refine impersonation attempts. We mitigate these risks by focusing on evaluation and guardrailing rather than providing actionable instructions for abuse, and by emphasizing deployment-oriented safeguards (e.g., configurable rules, conservative thresholds, and auditability). We encourage future work to further study fairness and privacy implications of speaker-attribute detection, and to adopt responsible release practices for audio datasets and models.

\section*{Acknowledgement}
This work is partially supported by the National Science Foundation under grant No. 1910100, No. 2046726, NSF AI Institute ACTION No. IIS-2229876, DARPA TIAMAT No. 80321, the National Aeronautics and Space Administration (NASA) under grant No. 80NSSC20M0229, ARL Grant W911NF-23-2-0137, Alfred P. Sloan Fellowship, the research grant from eBay, AI Safety Fund, Virtue AI, and Schmidt Science.

\bibliography{main}
\bibliographystyle{plainnat}

%%%%%%%%%%%%%%%%%%%%%%%%%%%%%%%%%%%%%%%%%%%%%%%%%%%%%%%%%%%%%%%%%%%%%%%%%%%%%%%
%%%%%%%%%%%%%%%%%%%%%%%%%%%%%%%%%%%%%%%%%%%%%%%%%%%%%%%%%%%%%%%%%%%%%%%%%%%%%%%
% APPENDIX
%%%%%%%%%%%%%%%%%%%%%%%%%%%%%%%%%%%%%%%%%%%%%%%%%%%%%%%%%%%%%%%%%%%%%%%%%%%%%%%
%%%%%%%%%%%%%%%%%%%%%%%%%%%%%%%%%%%%%%%%%%%%%%%%%%%%%%%%%%%%%%%%%%%%%%%%%%%%%%%
\newpage
\appendix
\onecolumn
% \section{LLM prompts}
% \label{app:prompts}

% \section{Dataset Development Details}
% \label{app:detail}

% \resizebox{\textwidth}{!}{
% \begin{prompt}[title=xxx]
% \scriptsize{xxx}
% \end{prompt}}

\section{Additional Evaluation Results}
\label{app:exp}

\begin{table*}[t]
\centering
\caption{\textbf{Content guardrail accuracy across benchmarks.}
We evaluate the content guardrail accuracy (i.e., correctness of the ground-truth safe/unsafe tag for the audio content) of \textit{TextGuard} in \method and other baselines. For \dataset, we evaluate on the \textbf{Speech} split and the \textbf{ElevenLabs Red-Teaming} split (non-speech slices are excluded since they are not observable from transcripts).}
\label{tab:text_guardrail_comparison}
\resizebox{\textwidth}{!}{%
\begin{tabular}{l|cc|cccc}
\toprule
\multirow{4}{*}{\textbf{Method}} &
\multicolumn{2}{c|}{\textbf{\dataset}} &
\multirow{4}{*}{\makecell{\textbf{Nemotron-}\\\textbf{Audio}}} &
\multirow{4}{*}{\makecell{\textbf{Jailbreak-}\\\textbf{AudioBench}}} &
\multirow{4}{*}{\makecell{\textbf{Omni-}\\\textbf{SafetyBench}}} &
\multirow{4}{*}{\makecell{\textbf{AdvWave}}}\\
\cmidrule(lr){2-3}
& \makecell{\textbf{Speech}} &
\makecell{\textbf{ElevenLabs}\\\textbf{Red-Teaming Test}} & & & & \\
\midrule
Gemma 3n         & 0.801 & 0.828 & 0.614 & 0.723 & 0.960 & 0.834 \\
Qwen3-Omni       & 0.824 & 0.623 & 0.734 & 0.623 & 0.523 & 0.803 \\
Audio Flamingo 3 & 0.523 & 0.345 & 0.834 & 0.663 & 0.563 & 0.322 \\
Gemini 3         & 0.934 & 0.902 & 0.723 & 0.902 & 0.849 & 1.000 \\
GPT-Audio        & 0.846 & 0.876 & 0.605 & 0.903 & 0.916 & 0.988 \\
\midrule
TextGuard (Ours) & \textbf{0.936} & \textbf{0.959} & \textbf{0.862} & 0.895 & \textbf{0.849} & \textbf{1.000} \\
\bottomrule
\end{tabular}%
}
\end{table*}

\paragraph{Analysis of content guardrail performance.}
As shown in \cref{tab:text_guardrail_comparison}, \textit{TextGuard} delivers the strongest and most consistent transcript-level moderation performance overall. It achieves the best accuracy on the two \dataset splits and on Nemotron-Audio, reaching 0.936 on Speech, 0.959 on ElevenLabs Red-Teaming, and 0.862 on Nemotron-Audio. On Jailbreak-AudioBench, Omni-SafetyBench, and AdvWave, \textit{TextGuard} remains competitive with the strongest audio-LLM baselines, even when it is not uniquely the top system. This pattern suggests that the main advantage of \textit{TextGuard} is not simply better in-domain fitting, but stronger alignment to the policy-grounded label space and greater robustness to transcript imperfections.

\begin{table}[t]
\centering
\caption{\textbf{Sound guardrail accuracy on \dataset.}
We report the audio-native (waveform-only) guardrail accuracy of \textit{SoundGuard} across \dataset splits: \textbf{Speech}, \textbf{Non-Speech}, and \textbf{ElevenLabs Red-Teaming}.}
\label{tab:sound_guardrail_comparison}
\begin{tabular}{lccc}
\toprule
\textbf{Model} & \textbf{Speech} & \textbf{Non-Speech} & \makecell{\textbf{ElevenLabs}\\\textbf{Red-Teaming}} \\
\midrule
Gemma 3n         & 0.194 & 0.362 & 0.427 \\
Qwen3-Omni       & 0.245 & 0.320 & 0.362 \\
Audio Flamingo 3 & 0.235 & 0.230 & 0.103 \\
Gemini 3         & 0.357 & 0.573 & 0.732 \\
GPT-Audio        & 0.289 & 0.452 & 0.400 \\
\midrule
SoundGuard (Ours) & \textbf{0.832} & \textbf{0.875} & \textbf{0.790} \\
\bottomrule
\end{tabular}
\end{table}

\paragraph{Analysis of sound guardrail performance.}
\Cref{tab:sound_guardrail_comparison} shows that \textit{SoundGuard} is the main driver of improvement on audio-native risks. It substantially outperforms all audio-LLM baselines on every \dataset split, with especially large margins on Speech and Non-Speech where correctly identifying voice attributes or harmful sound events is essential. These gains support our core design choice of separating waveform-level safety detection from semantic moderation, rather than relying on a single monolithic audio judge to implicitly recover both.

\begin{table*}[t]
\centering
\caption{\textbf{Multilingual F1 of TextGuard.} Per-language F1 on \dataset (17 languages), comparing Base Gemma-1B vs.\ our finetuned TextGuard.}
\label{tab:multilingual_f1}
\begin{tabular}{lcc | @{\hspace{14pt}} lcc}
\toprule
\textbf{Language} & \textbf{Base} & \textbf{TextGuard} &
\textbf{Language} & \textbf{Base} & \textbf{TextGuard} \\
\midrule
en    & 0.657 & 0.862 & nl    & 0.466 & 0.618 \\
es    & 0.523 & 0.689 & cs    & 0.447 & 0.595 \\
fr    & 0.547 & 0.692 & ar    & 0.431 & 0.584 \\
de    & 0.538 & 0.678 & zh-cn & 0.462 & 0.621 \\
it    & 0.512 & 0.653 & ja    & 0.428 & 0.573 \\
pt    & 0.505 & 0.641 & hu    & 0.439 & 0.589 \\
pl    & 0.493 & 0.627 & ko    & 0.436 & 0.587 \\
tr    & 0.471 & 0.612 & hi    & 0.415 & 0.562 \\
ru    & 0.458 & 0.603 &       &       &       \\
\bottomrule
\end{tabular}
\end{table*}

\paragraph{Analysis of multilingual transfer.}
The multilingual results in \cref{tab:multilingual_f1} show that \textit{TextGuard} consistently improves over the base Gemma-1B model in all 17 languages, with gains spanning both high-resource and lower-resource languages. The improvement is largest in English, but the gains remain strong and remarkably uniform across Romance, Germanic, Slavic, and Asian languages, suggesting that the learned safety decision boundary transfers well beyond the training language.

\begin{table}[t]
\centering
\caption{\textbf{Threshold sensitivity of end-to-end \method on \dataset.} We vary global compositional thresholds $(\tau_k,\tau_\ell)$ while keeping trained SoundGuard and TextGuard fixed.}
\label{tab:threshold_sensitivity}
\begin{tabular}{lcccc}
\toprule
\textbf{Threshold} & \textbf{Speech} & \textbf{Non-Speech} & \makecell{\textbf{ElevenLabs}\\\textbf{RT}} & \textbf{Avg} \\
\midrule
(0.1, 0.1) & 0.215 & 0.532 & 0.327 & 0.358 \\
(0.3, 0.3) & 0.803 & 0.863 & 0.803 & 0.823 \\
(0.5, 0.5) & \textbf{0.832} & \textbf{0.876} & \textbf{0.890} & \textbf{0.866} \\
(0.7, 0.7) & 0.743 & 0.852 & 0.831 & 0.809 \\
(0.9, 0.9) & 0.332 & 0.629 & 0.251 & 0.404 \\
\bottomrule
\end{tabular}
\end{table}

\paragraph{Analysis of threshold sensitivity.}
\Cref{tab:threshold_sensitivity} shows that the end-to-end behavior of \method is sensitive to the compositional thresholds, but also exhibits a clear and stable operating region. Very low thresholds lead to severe performance degradation, likely because the system becomes over-sensitive and triggers too many incorrect decisions. Conversely, very high thresholds also hurt performance, indicating that overly conservative gating suppresses true positives. The best results are achieved at the intermediate setting $(0.5, 0.5)$, which balances false positives and false negatives across all three in-domain splits.

\begin{table}[t]
\centering
\caption{\textbf{End-to-end \method accuracy with different SoundGuard encoders on \dataset.} ContentGuard and compositional integration remain fixed.}
\label{tab:stronger_encoder}
\begin{tabular}{lcccc}
\toprule
\textbf{SoundGuard backbone} & \textbf{Speech} & \textbf{Non-Speech} & \makecell{\textbf{ElevenLabs}\\\textbf{RT}} & \textbf{Avg} \\
\midrule
ECAPA-TDNN + MLP & 0.832 & \textbf{0.876} & \textbf{0.890} & \textbf{0.866} \\
WavLM-Large + MLP & \textbf{0.853} & 0.863 & 0.845 & 0.854 \\
\bottomrule
\end{tabular}
\end{table}

\paragraph{Analysis of SoundGuard backbones.}
The backbone comparison in \cref{tab:stronger_encoder} suggests that stronger encoders do not uniformly improve end-to-end guardrail quality. While WavLM-Large improves performance on the Speech split, ECAPA-TDNN yields better results on Non-Speech and ElevenLabs Red-Teaming and also achieves the best overall average. This indicates that end-to-end safety is not determined solely by encoder scale, but by how well the representation matches the mixture of speaker-aware and non-speech event detection required by the benchmark.

\begin{table}[t]
\centering
\caption{\textbf{Content-only accuracy of TextGuard on speech-based \dataset splits.}}
\label{tab:tts_asr_aug}
\begin{tabular}{lccc}
\toprule
\textbf{Setting} & \textbf{Speech} & \makecell{\textbf{ElevenLabs}\\\textbf{Content Acc.}} & \textbf{Avg} \\
\midrule
w/o TTS$\rightarrow$ASR augmentation & 0.902 & 0.932 & 0.917 \\
w/ TTS$\rightarrow$ASR augmentation & \textbf{0.936} & \textbf{0.959} & \textbf{0.948} \\
\bottomrule
\end{tabular}
\end{table}

\paragraph{Analysis of TTS$\rightarrow$ASR augmentation.}
\Cref{tab:tts_asr_aug} confirms that transcript-noise augmentation is important for robust semantic moderation. Adding the TTS$\rightarrow$ASR round-trip improves content accuracy on both the Speech split and the ElevenLabs split, with gains that are consistent across settings. This supports the hypothesis that clean-text supervision alone is insufficient for deployment conditions, where ASR artifacts can distort the phrasing of unsafe content or remove critical lexical cues.

\begin{table}[t]
\centering
\caption{\textbf{Robustness to noisy-overlap corruption on the \dataset Speech split.} We mix one of 10 background-noise wavs into the input audio with random overlap using \texttt{ffmpeg}.}
\label{tab:noisy_overlap}
\begin{tabular}{lccc}
\toprule
\textbf{Condition} & \textbf{Gemini 3} & \textbf{GPT-Audio} & \textbf{\method} \\
\midrule
Clean               & 0.357 & 0.289 & \textbf{0.832} \\
Noise / Overlapping & 0.355 & 0.274 & \textbf{0.793} \\
\bottomrule
\end{tabular}
\end{table}

\paragraph{Analysis of noisy-overlap robustness.}
\Cref{tab:noisy_overlap} shows that \method remains clearly strongest under realistic background-noise and overlap corruption. Although performance drops under corruption for all methods, the degradation for \method is modest relative to its clean-score advantage, suggesting that the modular design remains effective even when the audio is no longer cleanly segmented.

\paragraph{Benchmark quality and annotation.}
\dataset includes 13,535 English instances in the full benchmark. For the multilingual speech/content expansion, we translate 1,350 source examples into each of 16 additional languages, yielding 30,150 total multilingual speech/content instances. The English content subset contains 4,275 unsafe instances and 4,275 hard-benign instances. In the human verification stage, we use three annotators and retain only examples with unanimous agreement, yielding 100\% agreement on the retained set. These statistics make the benchmark scale, multilingual coverage, and false-positive evaluation protocol more transparent.

\paragraph{Latency protocol.}
All latency numbers in the main text are end-to-end wall-clock times. SoundGuard, ASR, TextGuard, and the full \method pipeline are measured locally on an NVIDIA A6000 GPU. Proprietary baselines such as Gemini 3 and GPT-Audio are measured via API end-to-end wall-clock time, which includes realistic transport overhead. We therefore interpret the latency comparison as deployment-oriented end-to-end latency rather than isolated model-only inference time.

\subsection{Baseline Prompting and Parsing}
\label{app:prompt}

For all end-to-end audio-LLM baselines, we use a fixed prompt template and a fixed two-line output schema. The goal is to minimize prompt variance across models and evaluate them under the same taxonomy and parsing protocol.

\begin{prompt}{Baseline prompt template}
You are an audio safety analyzer. Your task is to:

1. Output the identity of the speaker (Celebrity name, Child, or Unknown Speaker)

2. Classify the content safety

Content safety guideline:
Hate, Sexual, Self-Harm, Violence, Weapons, Privacy, Criminal, Harassment, Drugs, Illegal, Unauthorized Advice, Misinformation, Fraud, Terrorism, Other Risks

Please respond in EXACTLY this format:

Line 1: Speaker: [Celebrity name or "Child" or "Unknown Speaker"]

Line 2: Safety: [safe OR unsafe - Category]
\end{prompt}

We parse baseline outputs by exact string matching on this fixed two-line schema. For external benchmarks, when speaker identity is unspecified, we treat it as \emph{Unknown Speaker}. When an external benchmark does not align exactly with our 15-category taxonomy, we evaluate all methods consistently at the \textsc{safe} vs.\ \textsc{unsafe} level.

\subsection{Label Inventory}
\label{app:labels}

\paragraph{TextGuard semantic label space.}
TextGuard predicts one of the following semantic labels: \textsc{Safe}, Hate, Sexual, Self-Harm, Violence, Weapons, Privacy, Criminal, Harassment, Drugs, Illegal, Unauthorized Advice, Misinformation, Fraud, Terrorism, and Other Risks.

\paragraph{SoundGuard label-space schema.}
SoundGuard predicts waveform-level audio-native cues. Its label space contains two parts:
\begin{itemize}[leftmargin=1.5em,itemsep=0.2em]
    \item \textbf{Speaker-aware cues:} a large identity inventory consisting of 6k+ celebrity/public-figure identities sourced from VoxCeleb2 and supplemented with manually added missing public figures, plus a dedicated \textsc{Child} label.
    \item \textbf{Audio-native event cues:} harmful non-speech sound-event categories such as gunfire/explosion-like sounds, distress-related sounds, sexual sounds, breaking/forced-entry sounds, violence/struggle sounds, and crash-related sounds.
\end{itemize}

\end{document}